%% file: 00-main.tex
\documentclass[10pt,journal]{IEEEtran}
\newcommand{\para}[1]{\vspace{2pt}\noindent\textbf{#1.~}}
\newcommand{\system}{\sloppy{SmartMemory\@}}
\newcommand{\bug}{\sloppy{OFCI\@}}
\usepackage{diagbox}
\usepackage{enumitem}

\usepackage{pifont}
\usepackage{textcomp}
\usepackage{tabularx}
\usepackage{array}
\usepackage{ragged2e}
\usepackage{float}
\usepackage{listings, xcolor}
\usepackage{pifont}
\usepackage{ulem}
\usepackage{array,framed}
\usepackage{setspace}
\usepackage{url}

\expandafter\def\expandafter\UrlBreaks\expandafter{\UrlBreaks\do\.\do\@\do\;}
\usepackage{enumerate}
\usepackage{graphicx}
\usepackage{xspace}
\usepackage{multirow}
\usepackage{microtype}
\usepackage{amsmath}
\usepackage{amssymb}

\usepackage{listings, xcolor}
\definecolor{verylightgray}{rgb}{.97,.97,.97}
\lstdefinelanguage{Solidity}{
  keywords=[1]{anonymous, assembly, assert, balance, break, call, callcode, case, catch, class, constant, continue, constructor, contract, debugger, default, delegatecall, delete, do, else, emit, event, experimental, export, external, false, finally, for, function, gas, if, implements, import, in, indexed, instanceof, interface, internal, is, length, library, log0, log1, log2, log3, log4, memory, modifier, new, payable, pragma, private, protected, public, pure, push, require, return, returns, revert, selfdestruct, send, solidity, storage, struct, suicide, super, switch, then, this, throw, transfer, true, try, typeof, using, value, view, while, with, addmod, ecrecover, keccak256, mulmod, ripemd160, sha256, sha3}, 
  keywordstyle=[1]\color{blue}\bfseries,
  keywords=[2]{address, bool, byte, bytes, bytes1, bytes2, bytes3, bytes4, bytes5, bytes6, bytes7, bytes8, bytes9, bytes10, bytes11, bytes12, bytes13, bytes14, bytes15, bytes16, bytes17, bytes18, bytes19, bytes20, bytes21, bytes22, bytes23, bytes24, bytes25, bytes26, bytes27, bytes28, bytes29, bytes30, bytes31, bytes32, enum, int, int8, int16, int24, int32, int40, int48, int56, int64, int72, int80, int88, int96, int104, int112, int120, int128, int136, int144, int152, int160, int168, int176, int184, int192, int200, int208, int216, int224, int232, int240, int248, int256, mapping, string, uint, uint8, uint16, uint24, uint32, uint40, uint48, uint56, uint64, uint72, uint80, uint88, uint96, uint104, uint112, uint120, uint128, uint136, uint144, uint152, uint160, uint168, uint176, uint184, uint192, uint200, uint208, uint216, uint224, uint232, uint240, uint248, uint256, var, void, ether, finney, szabo, wei, days, hours, minutes, seconds, weeks, years},  
  keywordstyle=[2]\color{teal}\bfseries,
  keywords=[3]{block, blockhash, coinbase, difficulty, gaslimit, number, timestamp, msg, data, gas, sender, sig, value, now, tx, gasprice, origin},  
  keywordstyle=[3]\color{violet}\bfseries,
  identifierstyle=\color{black},
  sensitive=false,
  comment=[l]{//},
  morecomment=[s]{/*}{*/},
  commentstyle=\color{red}\ttfamily,
  stringstyle=\color{red}\ttfamily,
  morestring=[b]',
  morestring=[b]"
}
\begin{document}

\title{SmartMemory: Detecting On-chain-off-chain Communication Inconsistency for Smart Contract via Memory-based Agent}

\author{Zeqin~Liao,
        Yuhong~Nan,~\IEEEmembership{Member,~IEEE,}
        Henglong~Liang,
        Zixu~Gao,
		Lianyu~Hu$^{*}$,
        Yuqiang~Sun,
        Zhijie~Zhong,
        Xiaoyu~Ma,
        Zibin~Zheng,~\IEEEmembership{Fellow,~IEEE,}
        and~Yang~Liu,~\IEEEmembership{Senior~Member,~IEEE}%
\thanks{$^{*}$Corresponding author: Lianyu Hu.}%
\thanks{Zeqin Liao, Lianyu Hu, Yuqiang Sun, Xiaoyu Ma, and Yang Liu are with Nanyang Technological University, Singapore (e-mail: \protect\url{zeqin.liao@ntu.edu.sg}; \protect\url{hly2021@tju.edu.cn}; \protect\url{yuqiang.sun@ntu.edu.sg}; \protect\url{MA0009YU@e.ntu.edu.sg}; \protect\url{yangliu@ntu.edu.sg}).}%
\thanks{Yuhong Nan, Henglong Liang, Zixu Gao, Zhijie Zhong, and Zibin Zheng are with Sun Yat-sen University, China (e-mail: \protect\url{nanyh@mail.sysu.edu.cn}; \protect\url{lianghlong2@mail2.sysu.edu.cn}; \protect\url{zhongzh3@mail2.sysu.edu.cn}; \protect\url{gaozx9@mail2.sysu.edu.cn}; \protect\url{zhzibin@mail.sysu.edu.cn}).}}

\maketitle

\begin{abstract}


Smart contracts underpin decentralized finance, where growing demand for on-chain/off-chain communication (OFC) has driven diverse applications such as cross-chain bridges, real-world asset tokenization, and fiat-backed stablecoins. The OFC-related security incidents in these applications are increasingly frequent, but prior studies address separate vulnerability categories within OFC applications rather than providing a unified view, causing vulnerabilities outside known patterns to be missed.
%
In this paper, we identify \textit{OFC inconsistency} (OFCI) as a root cause of OFC vulnerabilities,
which arises from business-logic flaw and ultimately breaks the equivalence between the on-chain and off-chain asset representations to induce inconsistency.
%
Automatically detecting \bug{}s faces two challenges including (1) locating heterogeneous business logic, and (2) transferring existing vulnerability knowledge to identify unseen OFCI instances.

To this end, we propose \system{}, the first framework to leverage a memory-based agent for \bug{} detection. To address heterogeneity, SmartMemory maps diverse implementations of OFC contracts into a canonical business-semantic representation to locate the business logic for \bug{} inspection. For knowledge reuse, \system{} integrates a memory-based agent to distill vulnerability knowledge from features into patterns and detection rules, enabling knowledge transfer across cases to identify unseen \bug{}s. Lastly, \system{} performs taint analysis to verify the reachability, type, and impact of each candidate \bug{}.
We construct the first real-world \bug{} dataset comprising 48 DApps with 81 \bug{}s for evaluation, on which \system{} achieves 80.68\% precision and 87.65\% recall. In addition, through an analysis of 325 real-world OFC applications, \system{} detects 36 previously unknown \bug{}s, all of which have been confirmed and fixed by corresponding parties.

\end{abstract}


\input{01-body}

\bibliographystyle{IEEEtran}
\bibliography{reference}

\end{document}

%% file: 01-body.tex
\section{Introduction}
\label{sec:intro}

Smart contracts are programs deployed on blockchains, supporting a wide range of
decentralized applications (DApps) such as decentralized finance (DeFi)~\cite{chen2020defining}.  In the DeFi ecosystem, the demand for on-chain and off-chain communication (OFC) is rapidly growing, with typical applications such as cross-chain bridges~\cite{liao2024smartaxe}, real-world asset (RWA) tokenization~\cite{chen2024exploring}, and  fiat-backed stablecoins~\cite{guan2025security}.
However, security incidents targeting the on-chain contracts of these applications have become increasingly frequent~\cite{liao2024smartaxe, chen2024exploring}. Our investigation reveals more than 73 incidents in the past five years, the majority of which stem from contract vulnerabilities~\cite{statisticforOFCV}. 

Given its importance, the security of OFC contracts has not been systematically reviewed. More specifically, existing public reports and studies typically classify these incidents into seemingly-unrelated categories. For example, the vulnerabilities in PolyNetwork~\cite{PolyNetworkexploit}, Synapse~\cite{Synapse}, and THORChain~\cite{Thorchain} are attributed to an access-control vulnerability, accounting errors, and event forgery, respectively.
This fragmentation also appears at the analysis-tool level, where existing tools often address separate vulnerability categories within OFC applications rather than providing a unified view, and thus often miss vulnerabilities outside their patterns.
For example, SmartAxe~\cite{liao2024smartaxe}, BridgeGuard~\cite{zhou2025bridgeguard}, VCScope~\cite{wangpatterns}, and ChainSniper~\cite{tran2024chainsniper} focus on access-control and semantic vulnerabilities, interaction vulnerabilities, signature-verification vulnerabilities, and reentrancy or overflow vulnerabilities, respectively.

Despite the diverse categories and consequences, our research identifies  \textit{OFC inconsistency} (OFCI) as the common underlying cause of OFC vulnerabilities.
Specifically, OFC applications are expected to maintain value equivalence between the on-chain and off-chain representations of the same asset~\cite{ou2022overview}. We characterize this requirement as an \textit{equivalence invariant}.
While prior studies mainly target symptoms of a broken invariant, such as unauthorized state modification or accounting errors, we focus on where and how equivalence invariant is structurally broken.
Accordingly, we define \textit{OFC inconsistency}  as an inconsistency that arises from incomplete asset-exchange logic between on-chain contracts and off-chain entities and ultimately breaks the equivalence invariant.

While \bug{}s have been widely exploited in practice, research on OFC contract security remains limited.
First, vulnerability-detection tools~\cite{liao2024smartaxe, tran2024chainsniper, wangpatterns, zhou2025bridgeguard}
are dedicated to a specific OFC application domain (i.e., cross-chain bridges), and thus cannot cover general OFC exchange logic or generalize to other OFC applications such as RWA tokenization and fiat-backed stablecoins. 
Second, other related efforts~\cite{zhang2021dharcher, wang2024xguard, eshghie2024highguard} address adjacent problems rather than vulnerability detection. DArcher~\cite{zhang2021dharcher} tests data synchronization, while XGuard~\cite{wang2024xguard} and HighGuard~\cite{eshghie2024highguard} target behavior-anomaly analysis. 

Furthermore, existing approaches have limited generality in detecting \bug{}s due to two key limitations. 
First, most approaches~\cite{liao2024smartaxe,wangpatterns,tran2024chainsniper,zhou2025bridgeguard} rely on application-agnostic analysis or predefined rules, whereas \bug{}s are tightly coupled with application-specific business semantics. The same vulnerability may appear in different implementations of business logic, making it difficult for such analysis to capture.
Second, these approaches~\cite{liao2024smartaxe,tran2024chainsniper,zhou2025bridgeguard} are predominantly pattern-based and rely on human heuristics.
However, OFC applications involve $n$-to-$n$ combinations of chains, assets, and implementations, which causes their business logic and vulnerability patterns to grow combinatorially. As a result, any finite pattern set is insufficient in practice, exposing the limitation of pattern-based methods.
While the method~\cite{wangpatterns} based on a standalone LLM can reason over business logic for vulnerability discovery, it lacks a systematic mechanism to abstract, store, and transfer vulnerability knowledge across cases. This limits its ability to detect previously unseen  vulnerabilities.

Our insight is that a memory mechanism for LLM agent can address this limitation by enabling the agent to distill knowledge from detected cases and transfer it to new detection cases. In this way, the agent can accumulate and reuse knowledge across cases, helping it handle unseen vulnerabilities.

In this paper, we propose \system{}, an LLM-agent-based framework for detecting \bug{}s in smart contracts. 
\system{} operates through a three-stage pipeline of locating, identifying, and verifying \bug{}s.
In the first stage, \system{} uses a general conceptual model to map heterogeneous implementations into a canonical business-semantic representation, thereby locating the OFC business logic for inspection (Section~\ref{sec: semantic_extraction}).
In the second stage, \system{} uses a hierarchical-memory agent to check whether the target logic contains \bug{}. The memory progressively abstracts vulnerability knowledge from features into patterns and rules, enabling knowledge accumulation and transfer across cases (Section~\ref{sec: Inconsistency_Identification}). 
In the third stage, \system{} constructs a cross on-chain/off-chain data-flow graph (CDFG) and performs taint analysis on it to verify the reachability, type, and impact of the detected \bug{} candidate (Section~\ref{sec: Taint_Analysis}).

\para{Evaluation}
To evaluate the effectiveness of \system{}, we first construct a manually labeled dataset from public security reports of real-world OFC attacks, covering 48 OFC applications with 81 \bug{}s. Experimental results show that \system{} achieves 80.68\% precision and 87.65\% recall on the dataset. Compared with state-of-the-art (SOTA) tools within their respective detection scopes, \system{} significantly outperforms SmartAxe and VCScope. Furthermore, we construct a comprehensive OFC contract dataset consisting of 325 real-world OFC applications and deploy \system{} in collaboration with a professional auditing firm for large-scale security auditing. \system{} detects 36 previously unseen zero-day \bug{}s, all of which have been confirmed and fixed. 

\para{Contributions}
This paper makes the following contributions:

\begin{itemize}

\item 
We highlight \bug{} as a root cause of OFC vulnerabilities, which arises from business-logic flaw and ultimately breaks the equivalence invariants to induce inconsistency.


\item We propose \system{}, the first framework that leverages a memory-based agent to detect \bug{}s for OFC contracts, to the best of our knowledge.

\item We perform extensive experiments to demonstrate the effectiveness of \system{}. Through an analysis of 325 real-world applications, \system{} identifies 36 new \bug{}s, all of which have been confirmed and fixed.

\item We build the first manually-labeled dataset of OFC vulnerabilities, as well as the largest OFC contract dataset. 


\end{itemize}

\section{Background}
\label{sec:background}

\subsection{On-chain and Off-chain Communication in DeFi}
\label{sec: OFE}

\begin{figure} [t]
    \centering
    \includegraphics[width=1.05\linewidth]{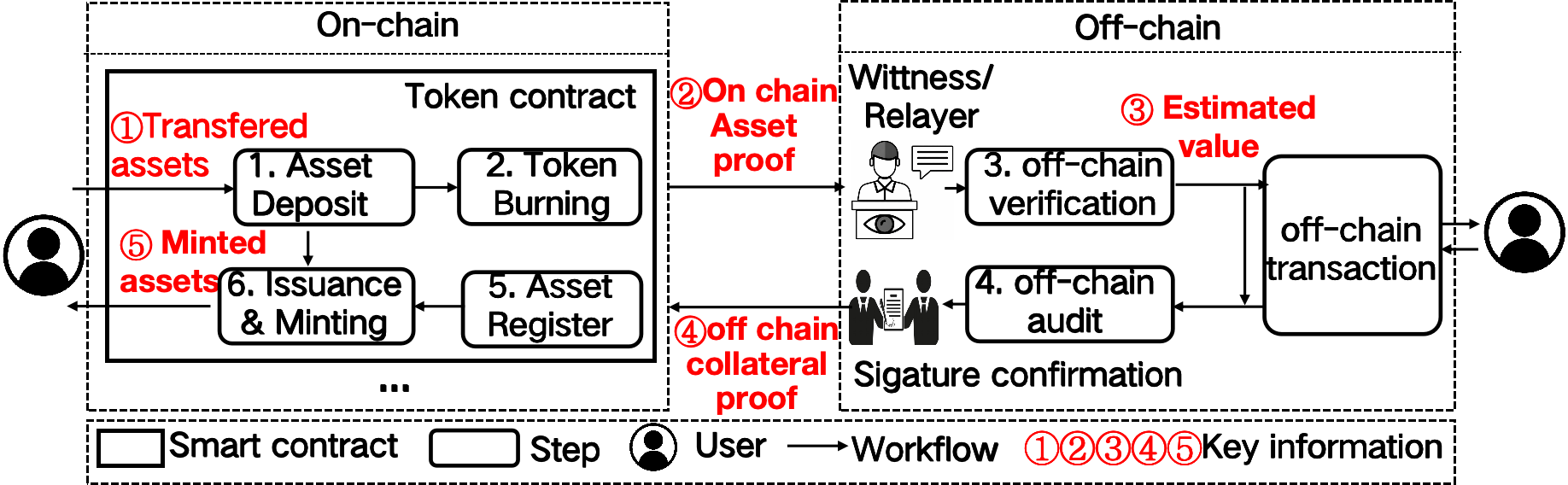}
    \caption{Core workflows of on-chain and off-chain communications}
    \label{fig:OFC}
    
\end{figure}

As a kind of smart-contract-powered peer-to-peer application, Decentralized Finance has attracted a surge in popularity~\cite{zhou2023sok}.

\para{On-chain and off-chain communication (OFC)}
In the DeFi ecosystem, on-chain and off-chain communication (OFC) refers to interactions between on-chain smart contracts and off-chain entities or external settlements~\cite{chen2024exploring, wangpatterns}. 
Fig.~\ref{fig:OFC}(a) shows an OFC architecture with on-chain and off-chain components. On-chain contracts implement token-exchange logic, while off-chain components relay information between contracts and entities. This allows users to deposit assets on one blockchain and redeem them off chain or on another blockchain, and vice versa. The OFC workflows are as follows:

\begin{itemize}[leftmargin=10pt]
    \item \textbf{Asset deposit.}
    The contract executes a deposit to transfer assets into the contract.

    \item \textbf{Token burning.}
    Upon redemption (see \textcircled{1}), the contract  burns the corresponding on-chain tokens, and emits an event as proof of asset outflow (see \textcircled{2}).

    \item \textbf{Off-chain verification.}
    Off-chain witnesses or relayers verify the on-chain proof and determine the asset value (see \textcircled{3}).

    \item \textbf{Off-chain audit.}
    Off-chain data are packaged as proof (see \textcircled{4}) and submitted to the contract.

    \item \textbf{Asset registration.}
    Upon receiving the proof, the contract verifies the submitted asset information for admission.

    \item \textbf{Issuance and minting.}
    Upon a minting request, the contract obtains the valuation, and mints tokens (see \textcircled{5}) under the issuance rules for the user.





\end{itemize}

\begin{figure*} [t]
    \centering
    \includegraphics[width=1\linewidth]{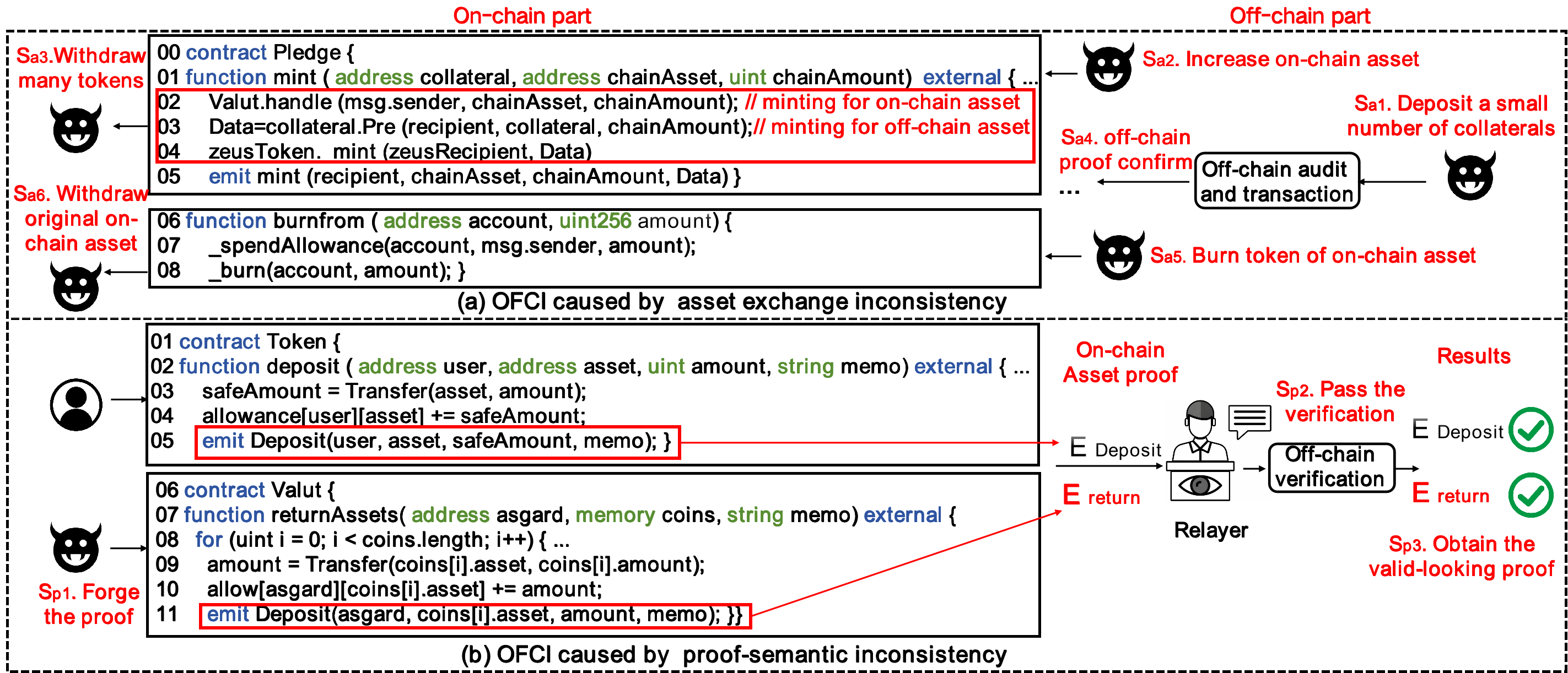}
    \caption{Examples of \bug{}s caused by different types of inconsistencies, namely, (a) asset-exchange inconsistency and (b) proof-semantic inconsistency.}
    \label{fig:motivating}
\end{figure*}

\para{Equivalence invariant}
OFC applications are required to maintain
value equivalence between the on-chain and off-chain representations of the same asset~\cite{han2023survey}. We characterize this requirement as equivalence invariant, i.e., the value represented on chain
must remain equivalent to its off-chain counterpart across each
relevant state transition of OFC applications.
Since the two representations
interact only through a small set of state transition points, the equivalence is constrained at
these seams~\cite{mao2022survey}. 
As shown in Fig. 1, OFC applications possess three equivalence invariants:

\begin{itemize}[leftmargin=10pt]
\item \textbf{$B_1$ (Outflow equivalence)}: the asset transferred by the user must be value-equivalent to the on-chain proof emitted by the contract (\textcircled{1}$\equiv$\textcircled{2}).

\item \textbf{$B_2$ (Verification equivalence)}: the on-chain proof emitted by the contract must be value-equivalent to the asset value parsed by the off-chain verifier (\textcircled{2}$\equiv$\textcircled{3}).

\item \textbf{$B_3$ (Issuance equivalence)}: the off-chain proof information accepted by the contract must be value-equivalent to the on-chain tokens minted from it (\textcircled{4}$\equiv$\textcircled{5}).
\end{itemize}

\subsection{LLM Agent and Memory}
\label{sec: LLM_Agent}

LLM-based agent is a rapidly developing research direction, which can continually learn and adapt to evolving task requirements~\cite{zhao2023depth,hu2025memory}.
Memory is a key mechanism supporting this capability ~\cite{zhang2025survey,hu2025memory,huang2026rethinking}. Analogous to human memory, it enables agents to learn from experience, reflect, and accumulate knowledge over time. It also distinguishes an agent from a standalone LLM~\cite{hu2025memory}.
Memory involves three core processes: (1) encoding, which transforms information into a storable form; (2) updating, which integrates encoded information into memory; and (3) retrieval, which extracts task-relevant information to prompt the model when needed~\cite{xie2024large}.
With memory, LLM agents can derive high-level insights from trajectories and experience. Produced through model introspection, these insights can be retrieved to improve current performance and support continual adaptation to dynamic task demands~\cite{hu2025memory,huang2026rethinking,xie2024large}.

\section{Motivation}
\label{sec: Motivation}

\subsection{Definition and Problem Statement}
\label{sec: ProblemStatement}




We define an \textit{OFC inconsistency} as a flaw in the contract's business logic that violates the equivalence invariants (i.e., $B_1$, $B_2$, and $B_3$ defined in Section~\ref{sec: OFE}), thereby causing the value of an asset's on-chain representation to diverge from its off-chain counterpart. 
We classify \bug{}s into two structural types based on which equivalence invariant is violated and where the resulting inconsistency is observed.
We use two motivating examples in Fig.~\ref{fig:motivating} to show two types of \bug{}s, which are drawn from real-world OFC applications exploited by attackers. We reorganize the original contract code for better clarity.

\textbf{Type-1: Asset-exchange inconsistency ($B_1$ and $B_3$).}
This type of inconsistency captures violations of asset-quantity equivalence in contract business logic, where the contract computes an output token quantity that is inconsistent with the corresponding input-side asset value. This type corresponds to violations of  $B_1$ and $B_3$.
Fig.~\ref{fig:motivating}(a) illustrates such a case caused by flawed business logic. In contract \texttt{Pledge}, function \texttt{mint} supports minting tokens according to either on-chain assets (line 2) or off-chain collateral (lines 3--4). However, both paths incorrectly use \texttt{chainAmount} to determine the mint amount, although this parameter actually denotes the amount derived from on-chain assets. Consequently, an adversary can submit a minting request with minimal off-chain collateral (see $S_{a1}$), then increase the on-chain asset amount (see $S_{a2}$), and obtain an excessive number of minted on-chain tokens (see $S_{a3}$). This creates an inconsistency between the off-chain proof (see $S_{a4}$) and the minted on-chain tokens (see $S_{a3}$). The adversary then withdraws the originally deposited on-chain assets (see $S_{a6}$), gaining profit with only minimal off-chain collateral.

\textbf{Type-2: Proof-semantic inconsistency ($B_2$).}
This type of inconsistency is caused by the flawed contract business logic for generating on-chain asset proofs, where the structure of an on-chain proof allows an off-chain verifier to parse a meaning inconsistent with the intended asset-flow semantics. This type corresponds to violations of $B_2$.
Fig.~\ref{fig:motivating}(b) illustrates such a case. In contract \texttt{Token}, function \texttt{deposit} locks on-chain assets (lines 3--4) and emits a Deposit event, $E_{deposit}$, as the proof of on-chain assets (line 5). However, another contract  \texttt{Vault} in the same DApp provides function \texttt{returnAssets} to batch-return assets from the vault to users. After returning the assets (lines 9--10), this function also emits a Transfer event, $E_{return}$ (line 11), with the same structure as $E_{deposit}$ . Moreover, each parameter of $E_{return}$ can be arbitrarily crafted through externally controlled inputs (line 11).
This design introduces an exploitable \bug{}. An adversary can invoke \texttt{returnAssets} to forge an on-chain asset proof (see $S_{p1}$), $E_{return}$. Because the off-chain relayer cannot distinguish $E_{return}$ from $E_{deposit}$, it may treat both as valid proofs (see $S_{p2}$), causing inconsistency between the on-chain proof and the verification result. As a result, the adversary may obtain a valid-looking asset proof (see $S_{p3}$) without actually transferring assets to the DApp, and then withdraw assets from the OFC application for profit.

\begin{figure}[t]
\small
\begin{lstlisting}
contract Wormhole {
function register(bytes memory encodedVm) { 
  VerifyVM(encodedVm); ... //P1: multi-sig verification  
  // P2: record message status (boolean flag)
  setTransferCompleted(vm.hash);  
  completedTransfers[hash] = true; ...
  // P3: emit event
  emit redeem(ChainId, Token, Recipient, Amount);}} 
contract Nomad {
function prove(bytes32 leaf, bytes32 proof, uint256 index) {
  // P1: verify message via Merkle proof + time window
  bytes32 root = MerkleLib.branch(leaf, proof, index);...
  // P2: record message status (Merkle root)
  messages[_hash] = root; ...  
  // P3: emit event
  emit Process(Hash, success, Data);
  emit Receive(nonce, token, recipient, amount);}}
\end{lstlisting}
\vspace{-1.8mm}
\caption{An example of diversity of off-chain asset registration.}
\label{diversityexample}
\vspace{-3mm}
\end{figure}

\para{Scope}
We target vulnerabilities where  controllable inputs cause the violation of $B_1$, $B_2$, or $B_3$ and induce on-chain/off-chain value inconsistency, which consists of  asset-exchange inconsistency and proof-semantic inconsistency. These two types cover the main OFC vulnerabilities, while fewer cases caused by non-contract issues (e.g., runtime/off-chain infrastructure) are outside the scope of our static analysis.

\subsection{Challenges and Solutions}
\label{sec: Challenges}

\bug{} originates from logic incompleteness (e.g., conflicting events or erroneous computations) and ultimately breaks equivalence invariant to induce inconsistency. Hence, a straightforward idea to identify \bug{}s has two key steps: (1) identify functions that implement incorrect or incomplete OFC-specific business logic and their relevant variables that an adversary can thereby manipulate; and (2) check whether critical variables involved in the equivalence invariant are data-flow dependent on manipulable variables in (1), so as to determine whether an adversary can indirectly alter them to induce profitable inconsistencies. 
However, implementing these steps is non-trivial.
We present the challenges and corresponding solutions.

\para{C1: Diversity in business-logic implementations} The first challenge is the diversity of OFC business-logic implementations, which stems from heterogeneous implementations and non-standardized designs~\cite{li2025towards}.
Fig.~\ref{diversityexample} shows two different implementations of off-chain asset registration in OFC contracts. \texttt{Wormhole}~\cite{Wormhole} verifies off-chain proofs through a guardian multisignature quorum (Line~3), records message status with a boolean flag (Lines~5--6), and emits \texttt{Redeem} (Line~8). 
In contrast, \texttt{Nomad}~\cite{Nomad} uses Merkle proofs (Line~12), stores the Merkle root as message status (Line~14), and emits \texttt{Process} and \texttt{Receive} (Lines~16--17).

Despite the implementation differences, OFC contracts are organized around core workflows (Section~\ref{sec: OFE}), which endow their business logic with a set of shared underlying semantics. We invite domain experts to distill these semantics into primitive functions, intra-operations, and operated variables, forming a general business-logic conceptual model for heterogeneous OFC contracts (Fig.~\ref{fig: conceptual}).
We further define four key variables around the equivalence breakpoints defined in Section~\ref{sec: ProblemStatement}, which capture OFC-specific on-chain/off-chain asset-exchange semantics that existing contract analysis techniques do not support.
With this model, \system{} prompts the LLM to assign corresponding business-logic-related statements with the triple labels of primitive function, intra-operation, and operated variable.
Then, \system{} produces semantically-annotated source code to locate the OFC business logic for inspection (Section~\ref{sec: semantic_extraction}).

\para{C2: Constructing memory mechanism for inconsistency detection} 
This challenge lies in the following aspects: (1) knowledge representation, where memory must encode information that can support \bug{} detection rather than store case details indiscriminately, (2) knowledge update and learning, where memory must distill reusable and generalizable \bug{} knowledge from successive detection cases rather than merely accumulate isolated observations, and (3) knowledge retrieval, where the relevant knowledge must be retrieved from a growing memory accurately for a given new \bug{} detection.

To address these challenges, we design an LLM agent with a three-layer memory that organizes knowledge into feature, pattern, and rule graphs.
For representation, the feature graph encodes \bug{}-specific information for individual vulnerabilities, while the pattern and rule graphs encode higher-level knowledge abstracted across them. For update, operators with explicit triggering conditions progressively abstract features into patterns and rules, merge redundant knowledge, remove outdated or falsified knowledge, and decompose overfitted knowledge, enabling the agent to abstract, store, and transfer vulnerability knowledge across cases. For retrieval, the agent augments similarity-based retrieval with graph-structural expansion, considering both semantic and structural relevance to reduce noisy matches
(Section~\ref{sec: Inconsistency_Identification}).


%
%


\section{Design of \system{}}
\label{sec: Approach}

\subsection{Overview}
\label{sec: Overview}

\begin{figure} [t]
    \centering
    \includegraphics[width=1\linewidth]{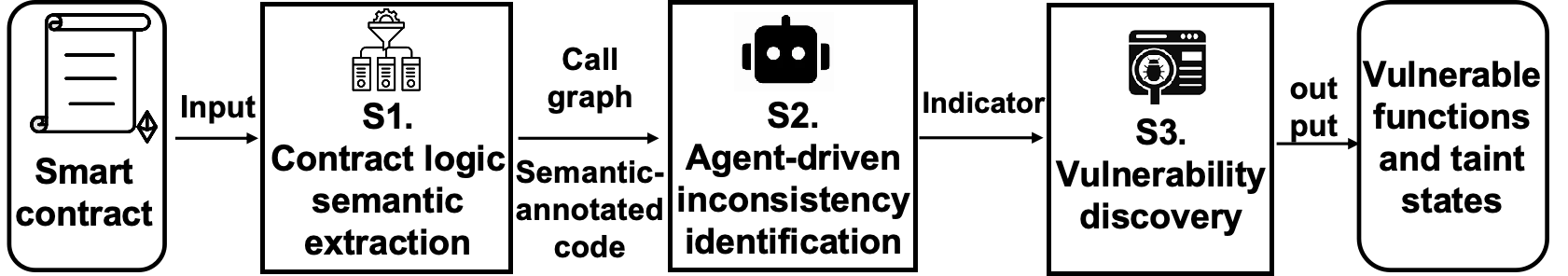}
    \caption{Overview of \system{}.}
    \label{fig:overview}
\vspace{-0.1 in}
\end{figure}

\system{} analyzes OFC smart contract source code and reports detected \bug{}s along with their vulnerability traces. As shown in Fig.~\ref{fig:overview}, the workflow forms a three-stage pipeline. The first locates the OFC business logic for inspection, the second identifies business-logic incompleteness, and the third verifies the reachability, type, and impact of detected \bug{}s.

\para{Contract-logic semantic extraction} Guided by the OFC conceptual model in Fig.~\ref{fig: conceptual}, \system{} identifies business-logic operations in the input contracts and annotates relevant statements with triple labels of primitive functions, intra-operations, and operated variables. With these annotations, \system{} generates semantically annotated source code to locate the OFC business logic for inspection and delimits the detection context with the constructed call graph.

\para{Agent-based inconsistency identification} \system{} employs an LLM agent with three-layer memory to examine whether the annotated target functions contain \bug{}s. This memory learns from an external vulnerability knowledge base and the agent's historical detection experience, abstracting low-level features into recurring patterns and generalized detection rules. For each target function, the agent retrieves relevant memory knowledge and combines it with the semantically-annotated source code to  prompt the LLM in determining whether the function is vulnerable.

\para{Vulnerability discovery via taint analysis} \system{} performs taint analysis on a cross on-chain/off-chain data-flow graph (CDFG). The analysis verifies whether an identified vulnerable function is externally reachable, whether taint can propagate to variables of equivalence invariant associated with possible inconsistencies, and determines the \bug{} type.

\subsection{Contract Logical Semantic Extraction}
\label{sec: semantic_extraction}


\para{Conceptual model construction for OFC contracts}
To build a generic business-logic model for heterogeneous OFC smart contracts, we first collect all the OFC applications from DeFiLlama~\cite{DeFiLlama}. We then invite domain experts to review the contracts in these DApps and identify relevant functions.
The experts first categorize the functions by OFC workflow, then abstract the primitive intra-operations in each category, and summarize the semantics of the key variables in each operation.

Fig.~\ref{fig: conceptual} presents the conceptual model of OFC contracts. The model is hierarchical: contracts contain functions, functions comprise intra-operations, and intra-operations operate over variables. 
The conceptual model consists of a set of primitive functions 
$\mathcal{O}=\{\mathit{Deposit},\allowbreak \mathit{Burn},\allowbreak \mathit{Register},\allowbreak \mathit{Mint},\allowbreak \mathit{Price},\allowbreak \mathit{Transfer},\allowbreak \mathit{Yield},\allowbreak \mathit{Govern},\allowbreak \mathit{Liquidate}\}$.
These nine primitive functions are systematically derived, rather than selected ad hoc, from the generic OFC workflows in Section~\ref{sec:background}. The four OFC-specific primitives (i.e., $\mathit{Deposit},\allowbreak \mathit{Burn},\allowbreak \mathit{Register},\allowbreak \mathit{Mint}$) abstract the core asset-exchange workflows, while the five general primitives (i.e., $\mathit{Price},\allowbreak \mathit{Transfer},\allowbreak \mathit{Yield},\allowbreak \mathit{Govern},\allowbreak \mathit{Liquidate}$) are included because they can become carriers that break the equivalence invariant.
For instance, the transfer operation \texttt{returnAssets} induces an inconsistency in Fig.~\ref{fig:motivating} (b).
Further, each function is composed of a certain subset of primitive intra-operations
$\mathcal{S}=\{\mathit{Check},\allowbreak \mathit{Update},\allowbreak \mathit{Emit},\allowbreak \mathit{Verify},\allowbreak \mathit{Assign}\}$ and
each intra operation operates on a certain subset of variables
$\mathcal{V}=\{\mathit{OnchainAsset},\allowbreak \mathit{OffchainCollateral},\allowbreak \mathit{Proof},\allowbreak \mathit{Debt},\allowbreak \mathit{Supply},\allowbreak \mathit{Valuation},\allowbreak \mathit{Permission},\allowbreak \mathit{Config},\allowbreak \mathit{Event}\}$. Further, we define the variables with 
$\mathcal{V}_{key}\in \{\mathit{OnchainAsset},\allowbreak \mathit{OffchainCollateral},\allowbreak \mathit{AssetProof}, \allowbreak \mathit{CollateralProof} \}$ as key variables, which correspond to the critical OFC variables at the two ends of the equivalence breakpoints.
Next, we introduce the composition of each primitive function in conceptual model.

\input{Figure/conceptualModel}

\textit{Deposit} validates asset admission with $\langle$Check, onchain-asset$\rangle$, updates the user balance and system supply with $\langle$Update, balance$\rangle$ and $\langle$Update, supply$\rangle$, and emits an on-chain inflow proof with $\langle$Emit, proof$\rangle$.

\textit{Burn} checks balance sufficiency with $\langle$Check, balance$\rangle$, deducts balance, debt, and total supply with $\langle$Update, balance$\rangle$, $\langle$Update, debt$\rangle$, and $\langle$Update, supply$\rangle$, and emits an asset-outflow proof with $\langle$Emit, proof$\rangle$.

\textit{Register} confirms that the asset is registered with $\langle$Check, onchain-asset$\rangle$, verifies the off-chain audit proof with $\langle$Verify, collateralproof$\rangle$, sets admission with $\langle$Update, onchain-asset$\rangle$, emits registration event with $\langle$Emit, event$\rangle$.

\textit{Mint} checks call compliance with $\langle$Check, permission$\rangle$ and validates the minting basis with $\langle$Check, onchain-asset$\rangle$ or $\langle$Check, offchain-collateral$\rangle$, allocates the minted amount with $\langle$Assign, onchain-asset$\rangle$, updates debt with $\langle$Update, debt$\rangle$, and emits a minting event with $\langle$Emit, event$\rangle$.

\textit{Price} restricts the caller to the oracle with $\langle$Check, permission$\rangle$, updates valuation with $\langle$Update, valuation$\rangle$, and emits a price-update event with $\langle$Emit, event$\rangle$.

\textit{Transfer} checks permission and balance with $\langle$Check, permission$\rangle$ and $\langle$Check, balance$\rangle$, updates both balances with $\langle$Update, balance$\rangle$, and emits transfer event $\langle$Emit, event$\rangle$.

\textit{Yield} credits computed yield to the user balance with $\langle$Update, balance$\rangle$ and emits a distribution event with $\langle$Emit, event$\rangle$.

\textit{Govern} updates risk parameters and system configuration, including contract upgrades, with $\langle$Update, valuation$\rangle$ and $\langle$Update, config$\rangle$, without a fixed order.

\textit{Liquidate} checks collateral insufficiency with $\langle$Check, debt$\rangle$, adjusts total supply and clears debt with $\langle$Update, supply$\rangle$ and $\langle$Update, debt$\rangle$, and emits liquidation event $\langle$Emit, event$\rangle$.

Note that the conceptual model is used to provide prototypes to construct structured prompts that guides the LLM in identifying contract logic for inspection.
Moreover, the OFC specificity of the model lies in the key variables, i.e., $\{\allowbreak \text{OnchainAsset}, \allowbreak \text{OffchainCollateral}, \allowbreak \mathit{AssetProof}, \allowbreak \mathit{CollateralProof}\}$, and their corresponding equivalence breakpoints. These equivalences are meaningful only in on-chain/off-chain asset-exchange scenarios and are not addressed by general contract analysis.


\para{Semantic annotation}
\system{} annotates the corresponding statement with a triple label $\langle o,\kappa,\delta\rangle$ to generate semantically-annotated code, where $o\in\mathcal{O}$, $\kappa\in \mathcal{S}$, and $\delta\in \mathcal{V}$. \system{} uses an LLM for semantic identification and annotates source code with triple labels. Given a function's source code and the conceptual-model prompt, the LLM identifies the $\langle o,\kappa,\delta\rangle$ labels corresponding to source statement. 
The labels are attached to statements, and multiple labels on the same statement are treated as a union.

Since annotation operates on statement semantics rather than parameter signatures, heterogeneous implementations of the same primitive operation can be mapped to the same label. Taking Fig.~\ref{diversityexample} as an example, \texttt{Wormhole} and \texttt{Nomad}  differ in parameter names, parameter counts, and implementation mechanisms, but can be annotated with the same triple $\langle$ Register, Update, onchain-asset$\rangle$. This is the core mechanism by which \system{} represents business-logic variants, i.e., implementation diversity is absorbed by semantic roles. \system{} does not replace real functions with templates from the conceptual model. The source code remains intact, and the model only guides the LLM in identifying semantics.

\system{} further constructs call graph, and then selects context functions along the call chain and those sharing the same triple label $\langle o,\kappa,\delta\rangle$, and serializes them into semantically-annotated code context. 
Here, the call graph provides the control-flow context (i.e., the call chain), while the triple label $\langle o,\kappa,\delta\rangle$ delimits the scope of the same business-logic operation.

\subsection{Agent-based Inconsistency Identification}
\label{sec: Inconsistency_Identification}


The memory design of \system{} is inspired by manual auditing practice, where experienced auditors extract case-specific features, abstract them into reusable patterns and rules, and update or retrieve such knowledge as new evidence accumulates.
Accordingly, \system{} structures its memory as three interconnected knowledge graphs including feature graph $G_{\text{feature}}$, pattern graph $G_{\text{pattern}}$, and rule graph $G_{\text{rule}}$.

Fig.~\ref{fig:InconsistencyDetection} illustrates the two-phase workflow of \system{}'s memory-based agent.

\textit{Phase-1. Memory retrieval for vulnerability detection}. Given the semantically annotated code context of a contract function, \system{} performs similarity-based top-down memory traversal to retrieve vulnerability features, patterns, and detection rules most relevant to the current detection task. \system{} then combines the retrieved knowledge with the code context to prompt the LLM to identify logic incompleteness.

\textit{Phase-2. Memory update for continual vulnerability learning}. After detection, \system{} extracts vulnerability features from the agent's historical detection experience and external vulnerability knowledge, and abstracts them into higher-level patterns and detection rules.

\begin{figure} [t]
    \centering
    \includegraphics[width=1.0\linewidth]{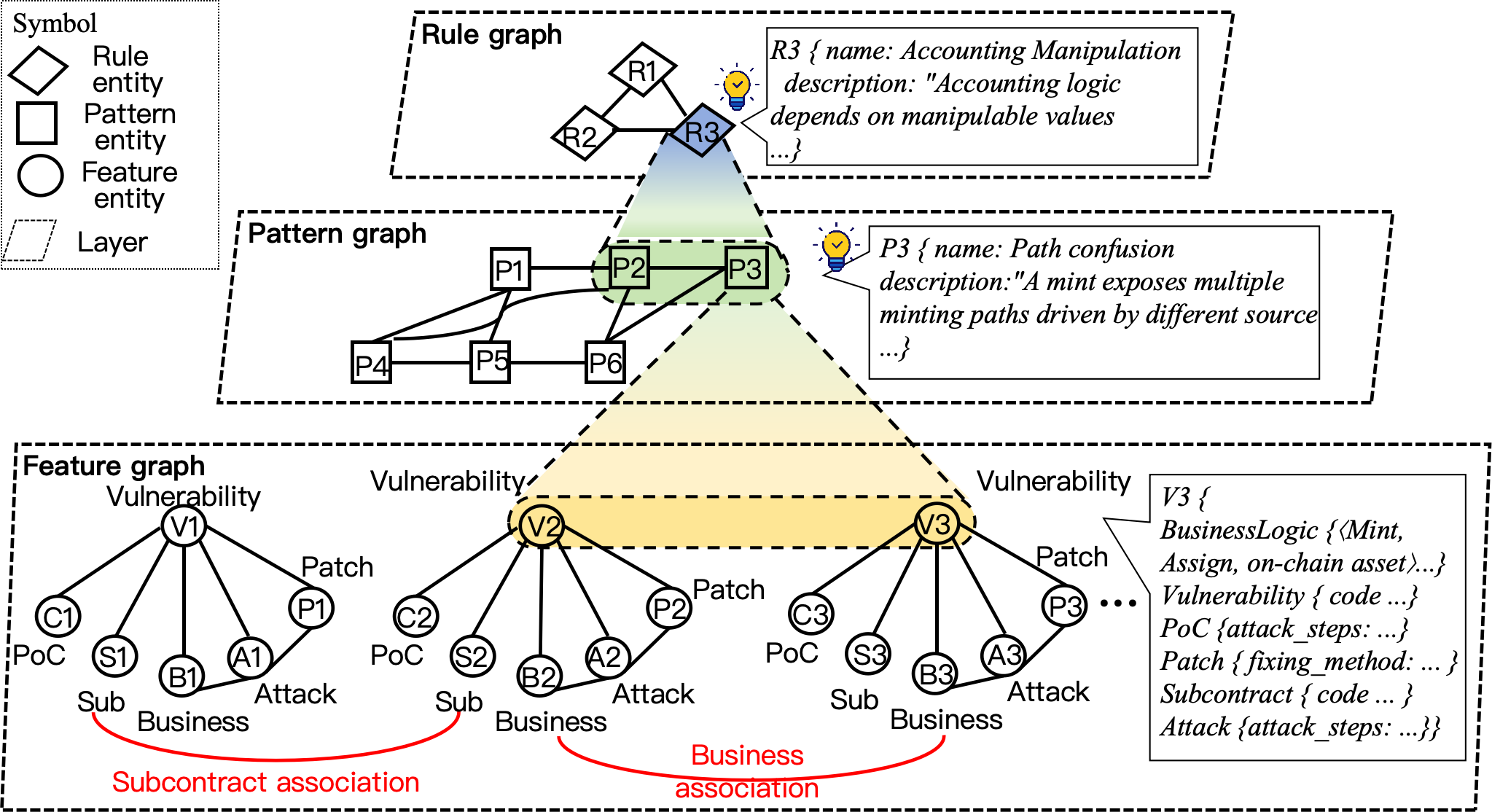}
    \caption{Architecture of three-layer memory integrated by \system{}.}
    \label{fig:3-memory}
    \vspace{-0.1 in}
    
\end{figure}

\begin{figure*} [t]
    \centering
    \includegraphics[width=1\linewidth]{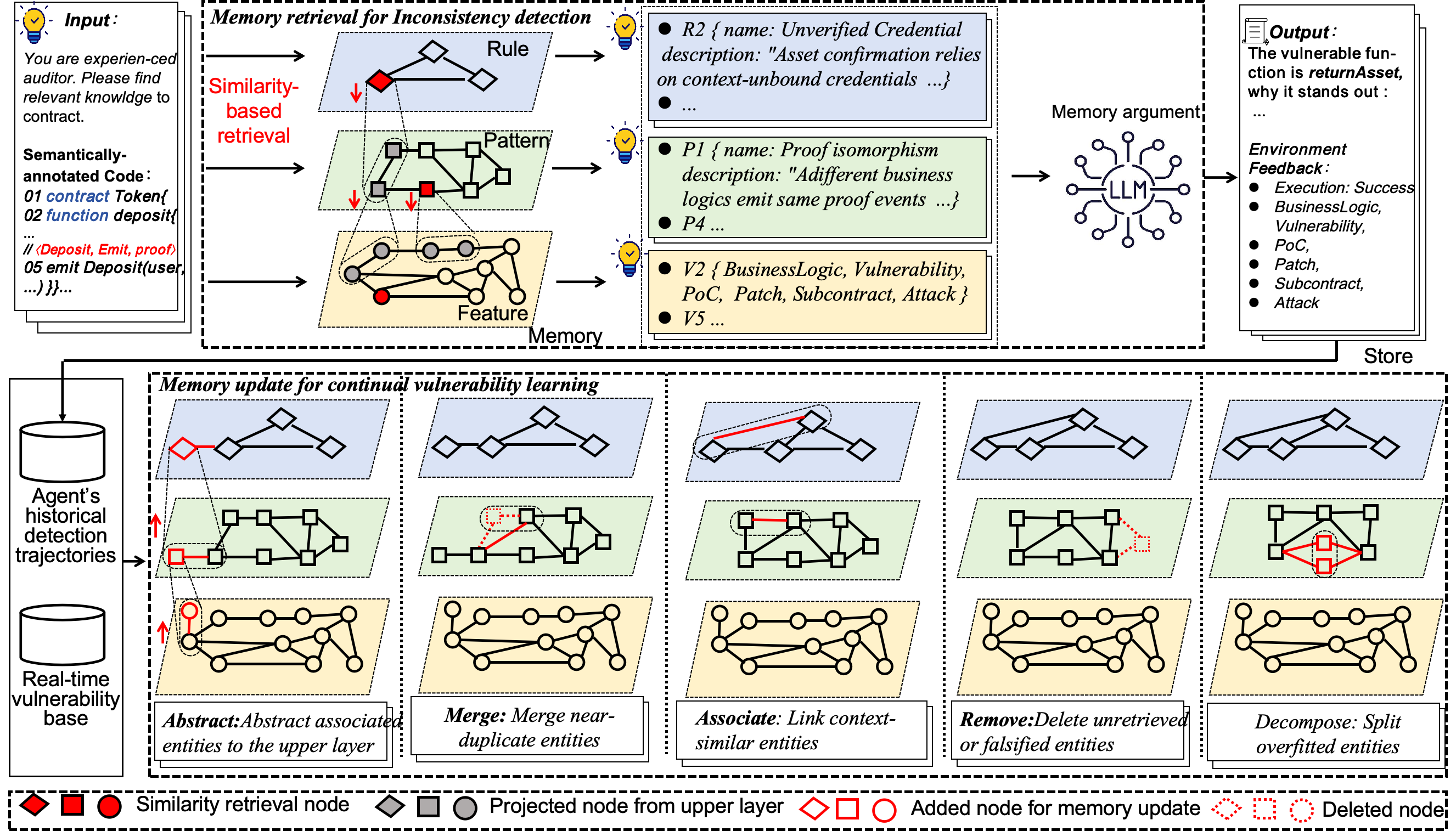}
    \caption{Details of inconsistency  identification in \system{}.}
    \label{fig:InconsistencyDetection}
    
\end{figure*}

\para{Three-layer memory structure} \system{} organizes its memory as a three-layer knowledge graph ordered by increasing levels of abstraction (Fig.~\ref{fig:3-memory}). Nodes within each layer encode vulnerability knowledge at the corresponding level, while inter-layer edges capture abstraction and derivation relations.

The feature layer captures structured knowledge of individual vulnerabilities. Each vulnerability is represented by six entity-node types: \textit{OFC Business Logic}, \textit{Vulnerability}, \textit{PoC}, \textit{Patch}, \textit{Subcontract}, and \textit{Attack}. Specifically, \textit{OFC Business Logic} describes the involved business logic using the conceptual-model triple $\langle o,\kappa,\delta\rangle$. \textit{Vulnerability} records the vulnerable code and a description of the inconsistency type and cause. \textit{PoC} records the proof-of-concept code and attack steps. \textit{Patch} records the patch code and fixing method. \textit{Subcontract} denotes involved subcontracts. And \textit{Attack} denotes the corresponding real-world attack event.
Within each vulnerability, these entities are linked by edges to form its feature subgraph. Specifically, a vulnerability \textit{occurs in} business logic, \textit{forms} a PoC, and \textit{locate at} a subcontract. An attack \textit{exploits} the vulnerability and \textit{manipulates} the business logic. A patch \textit{fixes} the vulnerability and \textit{blocks} the attack.
Feature subgraphs of different vulnerabilities are connected by association edges based on shared business logic, reused subcontracts, or shared attack events.

The pattern and rule layers contain natural-language nodes for vulnerability patterns and detection rules.
Within these two layers, nodes are connected by the three edge types inherited from the feature layer (i.e., shared business logic, reused subcontract, and shared attack event), as well as inter-entity association edges introduced during memory update.

Adjacent layers are linked by cross-layer edges that encode abstraction and derivation relations between upper and lower layer entities. These edges support top-down cross-layer projection during retrieval. 

\para{Memory retrieval}
\system{} traverses memory top-down, from the rule layer to the feature layer, as shown in Fig.~\ref{fig:InconsistencyDetection}.

\textit{Similarity-based retrieval.}
At each layer, \system{} retrieves the top-$k$ nodes most similar to a
query $q$, where similarity is measured by the cosine similarity between
the embeddings of the query and of each node's content. 
\begin{equation}
S = \operatorname{TopK}_{n_i \in N}
\frac{v(q)\cdot v(n_i)}{\|v(q)\|\,\|v(n_i)\|}.
\end{equation}
Since embedding similarity can be noisy, \system{} further performs a
one-hop expansion over the graph, adding the outgoing and incoming
neighbors of each node in $S$ to obtain $\tilde{S}$, which captures both
semantic and structural relevance.

\textit{Top-down memory traversal.}
\system{} retrieves layer by layer, from the rule layer down to the
feature layer. The query at each layer accumulates the knowledge
retrieved at the layers above it. Moreover, except for the top layer,
each layer first projects the results of the layer above onto itself
along the cross-layer edges to form seed nodes, and then merges these
seeds with its own retrieval results as the final results.

\noindent\textbf{(1) Rule layer.}
Using the targeted contract functions $C$ as the query, \system{} retrieves and expands
over $G_{\text{rule}}$ via Eq.~(1) to obtain the rule-layer
result $\tilde{R}_S$.

\noindent\textbf{(2) Pattern layer.}
\system{} first projects $\tilde{R}_S$ along the rule-to-pattern
cross-layer edges to the pattern nodes connected to them, forming a seed
set $P_{\text{proj}}$. Here, projection follows the cross-layer
edges to select adjacent nodes. \system{} then uses $[C;\,\tilde{R}_S]$ as the query to retrieve and expand over $G_{\text{pattern}}$, and
merges the result with $P_{\text{proj}}$ to obtain $\tilde{P}_S$.

\noindent\textbf{(3) Feature layer.}
Likewise, \system{} projects $\tilde{P}_S$ along the pattern-to-feature
cross-layer edges to form the feature seeds $F_{\text{proj}}$. \system{} then
uses $[C;\,\tilde{R}_S;\,\tilde{P}_S]$ as the query to retrieve and
expand over $G_{\text{feature}}$ via Eq.~(1), and merges the result
with $F_{\text{proj}}$ to obtain $\tilde{F}_S$.

\para{Memory learning} \system{} updates its memory using five operators, whose conditions and actions are as follows.

\begin{itemize} 
    \item \textit{Abstract} is triggered when a layer contains two or more entities connected by association edges, such as shared business logic, shared attack events, or shared subcontracts. \system{} utilizes an LLM to abstract these entities into a higher-level entity in the adjacent upper layer.
    \item \textit{Merge} merges two entities whose textual contents are semantically equivalent or highly similar according to their embedding similarity. 
    \item \textit{Associate} establishes an association edge between two entities whose neighborhood subgraphs are structurally similar, as measured by normalized graph edit distance~\cite{bunke1997relation}.
\item \textit{Remove} deletes an entity if it is persistently not retrieved in subsequent detections or if new evidence falsifies it or renders it outdated.
\item \textit{Decompose} utilizes an LLM to split a pattern-layer or rule-layer entity into finer-grained subentities when multiple detections show that it is overly case-specific, such as matching one case or failing to generalize to other variants.
\end{itemize}



Due to page limit,
the details of memory initialization, implementations of memory retrieval and update operators are available in our repository.

\system{} adapts memory-based agent to \bug{} detection through three domain-specific designs.
\textit{(i) Content.} The memory stores \bug{}-specific entities (e.g., OFC business-logic entities). \textit{(ii) Semantics.} The memory forms an audit-inspired abstraction hierarchy from features to patterns and rules. \textit{(iii) Operation.} The memory-update operators and top-down retrieval operator are designed to refine and transfer \bug{} knowledge.



\subsection{Vulnerability Discovery via Taint Analysis}
\label{sec: Taint_Analysis}

\system{} models the vulnerable functions identified above as \bug{} indicators and analyzes whether they can be triggered by an external adversary and whether they can affect critical OFC variables (e.g., \textcircled{1}--\textcircled{5} in equivalence invariant) to induce inconsistency. To this end, \system{} constructs a cross on-chain/off-chain data-flow graph (CDFG) and performs taint analysis to verify each indicator's reachability and impact, thereby determining the \bug{} type. 

\para{CDFG construction}
For an OFC contract, an asset proof must be confirmed off chain by a relayer or witness through verification before being consumed by subsequent logic. This off-chain confirmation introduces a data-flow break between proof production and proof consumption, which conventional data-flow analysis cannot continuously capture. To address this, \system{} first uses Slither~\cite{feist2019Slither} to construct control-flow graphs (CFGs) for OFC contracts and then performs data-flow analysis on the CFGs to obtain local data-flow graphs (DFGs). Among the critical variables, \textcircled{2}, \textcircled{3}, and \textcircled{4} in Fig.~\ref{fig:OFC} serve as alignment points across local DFGs. Since \textcircled{3} represents the asset value parsed off chain by the relayer or witness, \system{} does not analyze their implementation code and abstracts \textcircled{3} as a single node. \system{} then adds two alignment edges, one between \textcircled{2} and \textcircled{3}, and the other between \textcircled{3} and \textcircled{4}. These edges bridge proof production, off-chain confirmation, and proof submission, thereby connecting local DFGs across contracts into a unified CDFG.


\para{Taint analysis}
\system{} defines that sources are externally controllable public-function inputs while sinks are \bug{} indicators and OFC critical variables \textcircled{1}--\textcircled{5}. \system{} adopts the taint-propagation method from a prior study~\cite{liao2023smartstate} and confirms \bug{}s using two conditions.

\textbf{Condition 1: External-call reachability.}
On CDFG, \system{} performs forward taint propagation from the sources. If taint reaches an \bug{} indicator, the indicator is considered reachable, meaning that an external attacker can trigger it through controllable inputs.

\textbf{Condition 2: Critical-variable impact.}
Given a reachable indicator, \system{} checks whether taint can propagate to OFC critical variables involved in inconsistency checking.

\bug{} type depends on which critical-variable group the indicator affects. If the affected pair is \textcircled{1} and \textcircled{2}, or \textcircled{4} and \textcircled{5}, \system{} classifies the case as Type-1 asset-exchange inconsistency. If the affected pair is \textcircled{2} and \textcircled{3}, \system{} classifies it as Type-2 proof-semantic inconsistency.

We use the two cases in Fig.~\ref{fig:motivating} as running examples to illustrate how \system{} identifies the two \bug{} types.
In Fig.~\ref{fig:motivating}(a), \texttt{mint} contains two minting paths. During contract-logic semantic extraction, \system{} annotates the two paths as $\langle$Mint, Assign, onchain-asset$\rangle$ at Line~2 and $\langle$Mint, Assign, offchain-collateral$\rangle$ at Lines~3--4. During agent-based inconsistency identification, \system{} inspects Lines~2--4 and finds that both paths incorrectly reuse \texttt{chainAmount} as the source amount. \system{} identifies and marks \texttt{mint} as an \bug{} indicator. Taint analysis further verifies that \texttt{mint} can be triggered through a public entry and that taint can affect the off-chain collateral proof \textcircled{4} and the minted asset \textcircled{5}. \system{} classifies this case as asset-exchange inconsistency.
In Fig.~\ref{fig:motivating}(b), the event emitted by \texttt{Token.deposit} at Line~6 is annotated as $\langle$Deposit, Emit, proof$\rangle$, while the event emitted by \texttt{Vault.returnAssets} at Line~13 is annotated as $\langle$Transfer, Emit, event$\rangle$. \system{} inspects the two sites and finds that their emitted events are structurally isomorphic. \system{} therefore marks \texttt{returnAssets} as an \bug{} indicator. This proof isomorphism prevents the off-chain relayer or witness from distinguishing the forged proof from the genuine one. Taint analysis verifies that \texttt{returnAssets} is externally reachable and affects the produced proof \textcircled{2} and the parsed value \textcircled{3}. And \system{} classifies it as proof-semantic inconsistency.


\section{Evaluation}

\subsection{Implementation and Evaluation Setup}
\label{sec:setup}

\para{Research questions} The research questions are as follows:

\begin{itemize} 

\item RQ1. How effective is \system{} in detecting \bug{}s?

\item RQ2. How do the individual components of \system{} contribute to \bug{} detection?

\item RQ3. What are the runtime and monetary costs of \system{}?

\item RQ4. How effective is \system{} at finding zero-day vulnerabilities?

\end{itemize}

We use GPT-5-mini as the backbone LLM for \system{} during evaluation. 
All results are averaged over 5 runs on an Ubuntu 20.04 server with an Intel i9-10980XE CPU (3.0 GHz), an RTX 3090 GPU, and 250 GB RAM.

\para{Dataset and groundtruth establishment} We prepare two datasets and collection protocol (sources, search, exclusion, deduplication, and annotation schema) is available in repository.

\textit{Manually labeled \bug{} dataset} ($D_{labeled}$). To build ground truth for evaluating \system{}, we collect real-world \bug{} incidents from public security reports and disclosures, covering 48 distinct OFC applications (i.e., DApps). By reviewing the reports and contract code, we manually annotate 81 \bug{}s. 

\textit{Large-scale OFC contract dataset} ($D_{large}$). To evaluate  the real-world performance of \system{} on previously unseen \bug{}s, we exhaustively search the community and the Internet for OFC applications and collect 325 applications, comprising 5,158 code documents. 

We prevent information leakage through following isolation measures. First, the conceptual model contains only schema-level labels and encodes no application-specific instances. For $D_{\mathrm{labeled}}$, under leave-one-out reconstruction, 
we remove any information relevant to target DApp from vulnerability base and historical trajectories. The memory is rebuilt from the remaining N-1 applications. Third, for $D_{\mathrm{large}}$, we remove its overlap with $D_{\mathrm{labeled}}$ before evaluation, so the agent trained on $D_{\mathrm{labeled}}$ is tested only on contracts unseen in the labeled set. Together, these steps prevent target applications from entering memory construction or reappearing through labeled-set overlap.


\subsection{RQ1: Overall Effectiveness}
\label{sec:RQ1}

To answer RQ1, we evaluate \system{} on the manually labeled \bug{} dataset ($D_{labeled}$) using precision and recall. Following the leave-one-out reconstruction, we apply \system{} to each OFC application and manually compare its results with the ground truth (81 \bug{}s in 48 applications).
Here, we evaluate \system{}'s overall effectiveness in RQ1 and conduct fair comparisons with SOTA tools within their respective detection scopes in RQ2 below.

\input{Table/rq1_effectiveness.tex}

As shown in Table~\ref{tab:rq1_effectiveness}, \system{} achieves 80.68\% precision and 87.65\% recall on $D_{labeled}$, with consistently strong performance on both subtypes. 
The results indicate
that \system{} identifies \bug{}s for OFC contracts effectively.

\para{False positives and false negatives}
We inspect all 17 false positives and find that most of them originate from incomplete contract availability.
Some DApps do not open-source all deployed contracts, causing unresolved external calls during static analysis. 
Due to control-flow breaks, \system{} inevitably 
performs analysis with incomplete logic context, which increases false positives.
We also analyze the 10 false negatives and find that they mainly depend on non-contract components.
Among them, 4 rely on cryptographic verification logic in third-party libraries outside the OFC applications, and 6 depend on real-time data sources such as oracles.
%

\para{Temporal-holdout analysis} 
Since $D_{\text{labeled}}$ is from public security reports, some cases may have appeared in the backbone model's pre-training corpus and potentially introduce pre-training leakage.
To evaluate this risk, we split $D_{\text{labeled}}$ into pre-cutoff and post-cutoff subsets based on each \bug{}'s disclosure date and the model's knowledge cutoff date. 
Note that vulnerabilities disclosed after a model's knowledge cutoff are unlikely to have been included in its pre-training corpus.
We run SmartMemory on the full $D_{\mathrm{labeled}}$ using GPT-4o as the holdout backbone.
We use GPT-4o instead of GPT-5-mini because its knowledge cutoff (i.e., October 2023) is earlier and leaves enough post-cutoff \bug{}s for analysis.
%
We compare the precision and recall of \system{} on the two subsets. 

Among the 81 \bug{}s, 56 were disclosed before the cutoff date and 25 after it. With GPT-4o as the backbone, \system{} achieves 81.48\% (22/27) precision and 88.00\% (22/25) recall on the post-cutoff subset, comparable to the 80.00\% (48/60) precision and 85.71\% (48/56) recall on the pre-cutoff subset.
The results suggest that \system{} remains effective on \bug{}s without appearing in model pre-training, and its effectiveness is unlikely to be driven by pre-training leakage.
Notably, SmartMemory attains comparable precision and recall under both GPT-5-mini and GPT-4o, indicating that the method is model-agnostic and remains effective for different LLMs.

\subsection{RQ2: Ablation Study}
\label{sec:RQ2}

\para{Impact of contract-logic semantic extraction}
To answer RQ2, we evaluate contract-logic semantic extraction (Section~\ref{sec: semantic_extraction}), a core component of \system{}.
It maps diverse implementations to a common conceptual model, enabling business-logic identification across heterogeneous OFC contracts, and its impact is reflected in both precision and recall.
We compare the full \system{} with an ablation version without this module on $D_{labeled}$.

Table~\ref{tab:rq2_ablation} reports the results. Without this module, precision drops to 63.33\% and recall to 70.37\%, confirming that contract-logic semantic extraction substantially improves both metrics.
Moreover, 14 of 24 false negatives introduced by the ablation version can be resolved by our proposed method.
We revisit the motivating example in Fig.~\ref{fig:motivating}(a). Without semantic extraction, it cannot distinguish the two minting paths and therefore misses the conflicting use of \texttt{chainAmount}, causing a false negative.
In contrast, \system{} recognizes that the two paths correspond to different primitive operations with different expected parameters and eliminates the false negative.

\input{Table/rq2_ablation.tex}

\para{Impact of agent-driven inconsistency identification}
As discussed earlier, three-layer memory, top-down retrieval and  \bug{}-oriented update are three key designs of memory mechanism to support agent-driven inconsistency identification. We evaluate this component with three ablation versions.

We first implement a baseline version
of \system{} that removes all memory designs and relies on standalone LLM reasoning. 
We then implement multiple versions of \system{} by incrementally adding memory content, retrieval and update operators. We run all versions of \system{} on $D_{labeled}$, and further measure their precision and recall. 
 
As shown in Table~\ref{tab:rq2_ablation}, each design of memory mechanism contributes positively to the better
performance of \system{}, i.e., its precision and recall increase monotonically. Overall, the results indicate that memory mechanism helps \system{} enhance both precision and recall.

%

\para{Impact of taint-analysis-based vulnerability discovery}
Taint-analysis-based vulnerability discovery is another key component of \system{} to reveal the hidden \bug{}s, and improves the recall of \bug{} detection. ChainSniper and BridgeGuard are not open-sourced, XGuard, Darcher and HGuard do not support \bug{} detection. We conduct scope-matched comparisons with applicable SOTA tools.
We compare \system{} with SmartAxe~\cite{liao2024smartaxe} on access control and semantic vulnerabilities using 24 DApps from $D_{labeled}$, covering 41 \bug{} instances.
For VCScope~\cite{wangpatterns}, which targets signature-verification vulnerability, we select 9 DApps from $D_{labeled}$, covering 16 \bug{} instances.
We then compare recall on each evaluation set.

As shown in Table~\ref{tab:rq2_comparison}, \system{} achieves higher recall than both baselines within their respective scopes.
Against the 68.29\% recall of SmartAxe, \system{} achieves 87.80\% recall.
Against VCScope, \system{} reaches 87.50\% recall, compared with 43.75\%.
Further, manual investigation shows that our method can eliminate 8 of 13 false negatives from SmartAxe and 7 of 9 from VCScope.
%
The results indicate that taint-analysis-based vulnerability discovery helps \system{} outperform SOTA approaches.

\input{Table/rq2_comparison.tex}

\subsection{RQ3: Performance Overhead}
\label{sec:RQ3}

To answer RQ3, we evaluate the efficiency of \system{} on $D_{labeled}$ in terms of runtime and monetary cost.
Table~\ref{tab:rq3_efficiency} reports the average runtime and token usage per DApp for each stage. Contract-logic semantic extraction takes 110.66 seconds and 20,658 tokens (costing \$0.0072); agent-driven inconsistency identification takes 57.70 seconds and 25,492 tokens (costing \$0.0141); and vulnerability discovery takes 36.48 seconds. Overall, \system{} uses 204.84 seconds and 46,150 tokens per DApp on average, with an estimated cost of \$0.0213.
The results indicate that \system{} is practical in terms of runtime and cost. It can integrate locally deployed open-source LLMs to further reduce monetary cost.

\input{Table/rq3_efficiency.tex}

\subsection{RQ4: Finding Zero-days in the Real World}
\label{sec: Finding Zero-days}

To evaluate \system{}'s ability to discover previously unseen vulnerabilities, we apply it to $D_{large}$, which includes 325 real-world OFC DApps. This large-scale study is conducted with a professional security auditing firm that integrated \system{} into its auditing workflow.
\system{} reports 69 warnings (including 55 TPs and 14 FPs confirmed manually) across the 325 applications. 
After further manual inspection, our domain experts confirm 36 of the 55 TPs are exploitable zero-day \bug{}s.
The detailed inspection criteria are described in Section~\ref{sec:threat}.
For each confirmed vulnerability, we construct a proof-of-concept (PoC) exploit to verify exploitability. After responsible disclosure, we report all confirmed cases to the corresponding developers. As of submission, all of the 36 zero-day vulnerabilities have been acknowledged and fixed.
The disclosure materials are available in our repository.

%
%
%

\subsection{Threat to Validity}
\label{sec:threat}
One internal threat is subjective bias in manual analysis during dataset collection and evaluation. To mitigate it, we invited six domain experts and organized them into two annotator groups and one referee group. Each annotator had at least two years of experience, and each referee had at least four years of domain research experience. The annotator groups labeled independently and then cross-validated their results until consensus was reached. We used Cohen's Kappa to measure agreement. Cases below the threshold were escalated to the referee group for final adjudication. The average Cohen's Kappa across all pairs is 0.764, indicating substantial agreement.

\section{Related Work}
\label{sec:related}

\para{Smart contract vulnerability detection}
Research on the topic has evolved through several stages. Early work relies on program analysis, including symbolic execution~\cite{luu2016Making}, pattern matching~\cite{tikhomirov2018smartcheck,liao2022smartdagger,liao2025satellite}, and formal verification~\cite{tsankov2018securify}. Later studies use deep learning, such as attention-based BiLSTM~\cite{qian2020towards} and graph neural networks~\cite{zhuang2021smart}. More recently, LLM-based methods have emerged. While prior study~\cite{sun2024gptscan} combines LLMs with program analysis, other studies~\cite{wei2024ftsmartaudit, ma2025combining} fine-tune smaller LLMs or propose a two-stage framework of detection and explanation for auditing. However, these methods either rely on predefined patterns or lack continual learning from detection experience, limiting their adaptability to emerging OFC-specific vulnerability variants.

\para{Program analysis for OFC contract} Several approaches have been proposed for OFC contract security. SmartAxe~\cite{liao2024smartaxe} constructs access control models and cross-chain control-flow/data-flow graphs to detect cross-chain vulnerabilities. VCScope~\cite{wangpatterns} employs a two-stage LLM-based summarization method for vulnerability identification. ChainSniper~\cite{tran2024chainsniper} trains machine learning models on a limited dataset for cross-chain vulnerability detection. BridgeGuard~\cite{zhou2025bridgeguard} proposes a static analysis framework for detecting external interaction vulnerabilities.
Other notable efforts focus on anomaly detection for contract behavior.  
XGuard~\cite{wang2024xguard} applies predefined rules to detect inconsistent behaviors in cross-chain contracts. Hguard~\cite{eshghie2024highguard} uses manually constructed dynamic response graphs as formal specifications to identify logic deviations behavior.

\section{Conclusion}
\label{sec:conclusion}

In this paper, we propose SmartMemory, an agent-driven static-analysis framework to detect \bug{}s for OFC smart contracts.
On a manually labeled dataset of 81 \bug{}s from 48 OFC DApps, \system{} achieves 80.68\% precision and 87.65\% recall. We further apply \system{} to 325 real-world OFC DApps in collaboration with a professional security auditing firm, identifying 36 previously unseen zero-day \bug{}s, all of which have been acknowledged and patched. 

\section{Data Availability}
\label{sec:Data}


The repository (\url{https://doi.org/10.5281/zenodo.21006162}) provides the artifact, datasets, vulnerability statistics, complete prompts, parameter settings, embeddings, dataset collection protocol, memory implementations and disclosure materials. 


%% file: Figure/conceptualModel.tex
\begin{figure} [t]
  
\centering
\setlength{\fboxsep}{0pt}
\setlength{\fboxrule}{0.5pt}

\fbox{%
\begin{minipage}{\dimexpr\linewidth-2\fboxsep-2\fboxrule\relax}
\raggedright
\scriptsize
\renewcommand{\arraystretch}{1.08}
\noindent
$\displaystyle
\begin{array}{r@{\;::=\;}l}
\langle \mathit{OFCModel} \rangle
&
\mathit{C};f;\tau
\\[0pt]

\langle \mathit{Contract} \rangle
&
\mathit{C}=\{f_1,\ldots,f_n\}
\\[0pt]

\langle \mathit{Function} \rangle
&
f=\{s_1,\ldots,s_m\}
\\[0pt]

\langle \mathit{Operation} \rangle
&
\tau:S\rightarrow\langle \mathrm{o},\kappa,\delta \rangle
\\[0pt]


\langle \mathit{FunLabel \ \mathcal{O}} \rangle
&
\begin{array}[t]{@{}l@{}}
\mathrm{Deposit}\mid\mathrm{Burn}\mid\mathrm{Register}\mid\mathrm{Mint}
\mid\mathrm{Price}\\
{}\mid\mathrm{Transfer}\mid\mathrm{Yield}\mid\mathrm{Govern}
\mid\mathrm{Liquidate}
\end{array}
\\[0pt]

\langle \mathit{OpLabel \ \mathcal{K}} \rangle
&
\mathrm{Check}\mid\mathrm{Update}\mid\mathrm{Emit}\mid\mathrm{Verify}\mid\mathrm{Assign}
\\[0pt]

\langle \mathit{VarLabel \ \mathcal{V}} \rangle
&
\begin{array}[t]{@{}l@{}}
\mathrm{on\mbox{-}chain\ asset}\mid\mathrm{off\mbox{-}chain\ collateral}
\mid\mathrm{asset} \\ \mathrm{proof} \mid\mathrm{collateral\ proof}
{}\mid\mathrm{balance}\mid\mathrm{supply} \\ \mid\mathrm{debt}
\mid\mathrm{valuation}
{}\mid\mathrm{permission}\mid\mathrm{config}\mid\mathrm{event}
\end{array}
\\[0pt]

\langle \mathit{Composition} \rangle
&
\begin{array}[t]{@{}l@{}}
R(\mathrm{o})=\{(\kappa,\delta)\mid
\exists s\in \mathrm{S},
{}\tau(s)=\langle \mathrm{o},\kappa,\delta\rangle\}
\end{array}
\\[0pt]

\langle \mathit{Deposit} \rangle
&
\begin{array}[t]{@{}l@{}}
\langle \mathrm{Check},\mathrm{on\mbox{-}chain\ asset}\rangle,
\langle \mathrm{Update},\mathrm{balance}\rangle,\\
\langle \mathrm{Update},\mathrm{supply}\rangle,
\langle \mathrm{Emit},\mathrm{proof}\rangle
\end{array}
\\[0pt]

\langle \mathit{Burn} \rangle
&
\begin{array}[t]{@{}l@{}}
\langle \mathrm{Check},\mathrm{balance}\rangle,
\langle \mathrm{Update},\mathrm{balance}\rangle,\\
\langle \mathrm{Update},\mathrm{debt}\rangle,
\langle \mathrm{Update},\mathrm{supply}\rangle,
\langle \mathrm{Emit},\mathrm{proof}\rangle
\end{array}
\\[0pt]

\langle \mathit{Register} \rangle
&
\begin{array}[t]{@{}l@{}}
\langle \mathrm{Check},\mathrm{on\mbox{-}chain\ asset}\rangle,
\langle \mathrm{Verify},\mathrm{collateral\ proof}\rangle,\\
\langle \mathrm{Update},\mathrm{on\mbox{-}chain\ asset}\rangle,
\langle \mathrm{Emit},\mathrm{event}\rangle
\end{array}
\\[0pt]

\langle \mathit{Mint} \rangle
&
\begin{array}[t]{@{}l@{}}
\langle \mathrm{Check},\mathrm{permission}\rangle,
\langle \mathrm{Check},\mathrm{on\mbox{-}chain\ asset}\rangle,\\
\langle \mathrm{Check},\mathrm{off\mbox{-}chain\ collateral}\rangle,
\langle \mathrm{Update},\mathrm{debt}\rangle,\\
\langle \mathrm{Assign},\mathrm{on\mbox{-}chain\ asset}\rangle,
\langle \mathrm{Emit},\mathrm{event}\rangle
\end{array}
\\[0pt]

\langle \mathit{Price} \rangle
&
\begin{array}[t]{@{}l@{}}
\langle \mathrm{Check},\mathrm{permission}\rangle,
\langle \mathrm{Update},\mathrm{valuation}\rangle,\\
\langle \mathrm{Emit},\mathrm{event}\rangle
\end{array}
\\[0pt]

\langle \mathit{Transfer} \rangle
&
\begin{array}[t]{@{}l@{}}
\langle \mathrm{Check},\mathrm{permission}\rangle,
\langle \mathrm{Check},\mathrm{balance}\rangle,\\
\langle \mathrm{Update},\mathrm{balance}\rangle,
\langle \mathrm{Emit},\mathrm{event}\rangle
\end{array}
\\[0pt]

\langle \mathit{Yield} \rangle
&
\langle \mathrm{Update},\mathrm{balance}\rangle,
\langle \mathrm{Emit},\mathrm{event}\rangle
\\[0pt]

\langle \mathit{Govern} \rangle
&
\langle \mathrm{Update},\mathrm{valuation}\rangle,
\langle \mathrm{Update},\mathrm{config}\rangle
\\[0pt]

\langle \mathit{Liquidate} \rangle
&
\begin{array}[t]{@{}l@{}}
\langle \mathrm{Check},\mathrm{debt}\rangle,
\langle \mathrm{Update},\mathrm{debt}\rangle,
\langle \mathrm{Emit},\mathrm{event}\rangle
\end{array}
\end{array}
$
\end{minipage}%
}
\caption{ Conceptual model of OFC smart contracts }

\label{fig: conceptual}

\vspace{-0.1 in}

\end{figure}

%% file: Table/rq1_effectiveness.tex
\begin{table}[t]
\caption{Overall effectiveness of \system{} on $D_{labeled}$.}
\label{tab:rq1_effectiveness}
\centering
\footnotesize
\begin{tabular}{l|ccc|ccc}
\hline
\multirow{2}{*}{\bug{}} & \multicolumn{3}{c|}{Precision} & \multicolumn{3}{c}{Recall} \\
 & TP & FP & Rate & TP & FN & Rate \\
\hline
AEI &60  &14  &81.08\%  &60  &8  &88.24\%  \\
PI  &11  &3  &78.57\%  &11  &2  & 84.62\% \\
\hline
Total &71  &17  &80.68\%  &71  &10  &87.65\%  \\
\hline
\end{tabular}
\begin{flushleft}
\footnotesize
\textit{Note:} AEI: Asset-exchange inconsistency; PI: Proof-semantic Inconsistency.
\end{flushleft}
\vspace{-0.2 in}
\end{table}

%% file: Table/rq2_ablation.tex
\begin{table}[t]
\caption{Comparison results between \system{} and the ablation baselines over $D_{labeled}$.}
\label{tab:rq2_ablation}
\centering
\footnotesize
\setlength{\tabcolsep}{3pt}
\renewcommand{\arraystretch}{0.95}
\begin{tabular}{@{}c|l|ccc|ccc@{}}
\hline
\multirow{2}{*}{\shortstack{Component}} & \multirow{2}{*}{Approach} & \multicolumn{3}{c|}{Precision} & \multicolumn{3}{c}{Recall} \\
 & & TP & FP & Rate & TP & FN & Rate \\
\hline
\shortstack{Logic extraction} & w/o logic extraction       &57  &33  &63.33\%  &57  &24  &70.37\%  \\ \hline
\multirow{3}{*}{\shortstack{Inconsistency\\detection}} 
                  & Baseline (\textit{A})      &37  &25  &59.68\%     &37  &44  &45.68\%     \\ 
                  & \textit{A} + memory (\textit{B})     &51  &20  &71.83\%     &51  &30  &62.96\%     \\ 
                  & \textit{B} + retrieval ops (\textit{C})  &67  &24  &73.63\%     &67  &14  &82.72\%     \\ \hline
Full version & \begin{tabular}[c]{@{}l@{}}\system{}\\(\textit{C} + update ops)\end{tabular}              &71  &17  &80.68\%  &71  &10  &87.65\%  \\
\hline

\end{tabular}
\vspace{-0.1 in}
\end{table}

%% file: Table/rq2_comparison.tex
\begin{table}[t]
\caption{Scope-matched comparison between \system{} and SOTA tools within their respective scopes on $D_{labeled}$.}
\label{tab:rq2_comparison}
\centering
\footnotesize
\begin{tabular}{l|l|ccc}
\hline
\multirow{2}{*}{Scope} & \multirow{2}{*}{Approach} & \multicolumn{3}{c}{Recall} \\
 & & TP & FN & Rate \\
\hline
\multirow{2}{*}{\shortstack[l]{Cross-chain bridge\\(24 DApps, 41 \bug{}s)}}
 & SmartAxe~\cite{liao2024smartaxe} & 28 & 13 & 68.29\% \\
 & \system{}          & 36 & 5 & 87.80\% \\
\hline
\multirow{2}{*}{\shortstack[l]{Signature verification \\(9 DApps, 16 \bug{}s)}}
 & VCScope~\cite{wangpatterns}  & 7 & 9 & 43.75\% \\
 & \system{}          & 14 & 2 & 87.50\% \\
\hline
\end{tabular}
\end{table}

%% file: Table/rq3_efficiency.tex
\begin{table}[h]
\caption{The average time and tokens for analyzing each Dapp in $D_{labeled}$.}
\label{tab:rq3_efficiency}
\centering
\footnotesize
\setlength{\tabcolsep}{3pt}
\renewcommand{\arraystretch}{1.15}
\begin{tabularx}{\columnwidth}{@{}>{\raggedright\arraybackslash}X c c@{}}
\hline
Efficiency of \system{} & Avg. time (s) & Avg. tokens \\
\hline
Contract logical semantic extraction & 110.66 & 20658 (\$0.0072) \\
Agent-based inconsistency identification & 57.70 & 25492 (\$0.0141) \\
Vulnerability discovery & 36.48 & -- \\
\hline
Total & 204.84 & 46150 (\$0.0213) \\
\hline
\end{tabularx}
\end{table}

%% file: reference.bib
@misc{Wormhole,
  
  
  title={Wormhole Bridge Exploit Incident Analysis}, 
  howpublished = "\url{https://www.certik.com/blog/wormhole-bridge-exploit-incident-analysis}",
  note = "[Accessed 26-Mar-2026]",
}

@article{liao2025satellite,
  title={Satellite: Detecting and Analyzing Smart Contract Vulnerabilities caused by Subcontract Misuse},
  author={Liao, Zeqin and Nan, Yuhong and Gao, Zixu and Liang, Henglong and Hao, Sicheng and Wu, Jiajing and Zheng, Zibin},
  journal={IEEE Transactions on Software Engineering},
  year={2025},
  publisher={IEEE}
}

@article{bunke1997relation,
  title={On a relation between graph edit distance and maximum common subgraph},
  author={Bunke, Horst},
  journal={Pattern recognition letters},
  volume={18},
  number={8},
  pages={689--694},
  year={1997},
  publisher={Elsevier}
}

@misc{DeFiLlama,

  year = {2026},
  title={DeFiLlama}, 
  howpublished = "\url{https://defillama.com/}",
  note = "[Accessed 27-June-2026]",
}

@article{li2025towards,
  title={Towards blockchain interoperability: A comprehensive survey on cross-chain solutions},
  author={Li, Wenqing and Liu, Zhenguang and Chen, Jianhai and Liu, Zhe and He, Qinming},
  journal={Blockchain: Research and Applications},
  pages={100286},
  year={2025},
  publisher={Elsevier}
}

@article{han2023survey,
  title={A survey on cross-chain technologies},
  author={Han, Panpan and Yan, Zheng and Ding, Wenxiu and Fei, Shufan and Wan, Zhiguo},
  journal={Distributed ledger technologies: research and practice},
  volume={2},
  number={2},
  pages={1--30},
  year={2023},
  publisher={ACM New York, NY}
}

@inproceedings{liao2022smartdagger,
  title={SmartDagger: a bytecode-based static analysis approach for detecting cross-contract vulnerability},
  author={Liao, Zeqin and Zheng, Zibin and Chen, Xiao and Nan, Yuhong},
  booktitle={Proceedings of the 31st ACM SIGSOFT International Symposium on Software Testing and Analysis},
  pages={752--764},
  year={2022}
}

@inproceedings{zhou2023sok,
  title={Sok: Decentralized finance (defi) attacks},
  author={Zhou, Liyi and Xiong, Xihan and Ernstberger, Jens and Chaliasos, Stefanos and Wang, Zhipeng and Wang, Ye and Qin, Kaihua and Wattenhofer, Roger and Song, Dawn and Gervais, Arthur},
  booktitle={2023 IEEE Symposium on Security and Privacy (SP)},
  pages={2444--2461},
  year={2023},
  organization={IEEE}
}

@article{mao2022survey,
  title={A survey on cross-chain technology: Challenges, development, and prospect},
  author={Mao, Hanyu and Nie, Tiezheng and Sun, Hao and Shen, Derong and Yu, Ge},
  journal={IEEE Access},
  volume={11},
  pages={45527--45546},
  year={2022},
  publisher={IEEE}
}

@article{ou2022overview,
  title={An overview on cross-chain: Mechanism, platforms, challenges and advances},
  author={Ou, Wei and Huang, Shiying and Zheng, Jingjing and Zhang, Qionglu and Zeng, Guang and Han, Wenbao},
  journal={Computer Networks},
  volume={218},
  pages={109378},
  year={2022},
  publisher={Elsevier}
}

@inproceedings{zhang2021dharcher,
  title={Darcher: Detecting on-chain-off-chain synchronization bugs in decentralized applications},
  author={Zhang, Wuqi and Wei, Lili and Li, Shuqing and Liu, Yepang and Cheung, Shing-Chi},
  booktitle={Proceedings of the 29th ACM Joint Meeting on European Software Engineering Conference and Symposium on the Foundations of Software Engineering},
  pages={553--565},
  year={2021}
}

@inproceedings{ma2025combining,
  title={Combining fine-tuning and llm-based agents for intuitive smart contract auditing with justifications},
  author={Ma, Wei and Wu, Daoyuan and Sun, Yuqiang and Wang, Tianwen and Liu, Shangqing and Zhang, Jian and Xue, Yue and Liu, Yang},
  booktitle={2025 IEEE/ACM 47th International Conference on Software Engineering (ICSE)},
  pages={1742--1754},
  year={2025},
  organization={IEEE}
}

@article{zhou2025bridgeguard,
  title={BridgeGuard: Checking external interaction vulnerabilities in cross-chain bridge router contracts based on symbolic dataflow analysis},
  author={Zhou, Zequan and Luo, Xiling and Ji, Xiaohai and Mao, Jian and He, Ting and Wang, Junjun and Wu, Qianhong},
  journal={IEEE Transactions on Dependable and Secure Computing},
  year={2025},
  publisher={IEEE}
}

@article{wei2024ftsmartaudit,
  title={Ftsmartaudit: A knowledge distillation-enhanced framework for automated smart contract auditing using fine-tuned llms},
  author={Wei, Zhiyuan and Sun, Jing and Zhang, Zijian and Zhang, Xianhao and Hou, Zhe},
  journal={arXiv preprint arXiv:2410.13918},
  year={2024}
}

@inproceedings{zhuang2021smart,
  title={Smart contract vulnerability detection using graph neural networks},
  author={Zhuang, Yuan and Liu, Zhenguang and Qian, Peng and Liu, Qi and Wang, Xiang and He, Qinming},
  booktitle={Proceedings of the twenty-ninth international conference on international joint conferences on artificial intelligence},
  pages={3283--3290},
  year={2021}
}

@article{qian2020towards,
  title={Towards automated reentrancy detection for smart contracts based on sequential models},
  author={Qian, Peng and Liu, Zhenguang and He, Qinming and Zimmermann, Roger and Wang, Xun},
  journal={IEEE access},
  volume={8},
  pages={19685--19695},
  year={2020},
  publisher={IEEE}
}

@inproceedings{tsankov2018securify,
  title={Securify: Practical security analysis of smart contracts},
  author={Tsankov, Petar and Dan, Andrei and Drachsler-Cohen, Dana and Gervais, Arthur and Buenzli, Florian and Vechev, Martin},
  booktitle={Proceedings of the 2018 ACM SIGSAC conference on computer and communications security},
  pages={67--82},
  year={2018}
}

@inproceedings{tikhomirov2018smartcheck,
  title={Smartcheck: Static analysis of ethereum smart contracts},
  author={Tikhomirov, Sergei and Voskresenskaya, Ekaterina and Ivanitskiy, Ivan and Takhaviev, Ramil and Marchenko, Evgeny and Alexandrov, Yaroslav},
  booktitle={Proceedings of the 1st international workshop on emerging trends in software engineering for blockchain},
  pages={9--16},
  year={2018}
}

@inproceedings{wang2024xguard,
  title={Xguard: Detecting inconsistency behaviors of crosschain bridges},
  author={Wang, Ke and Li, Yue and Wang, Che and Gao, Jianbo and Guan, Zhi and Chen, Zhong},
  booktitle={Companion Proceedings of the 32nd ACM International Conference on the Foundations of Software Engineering},
  pages={612--616},
  year={2024}
}

@inproceedings{eshghie2024highguard,
  title={Highguard: Cross-chain business logic monitoring of smart contracts},
  author={Eshghie, Mojtaba and Artho, Cyrille and Stammler, Hans and Ahrendt, Wolfgang and Hildebrandt, Thomas and Schneider, Gerardo},
  booktitle={Proceedings of the 39th IEEE/ACM International Conference on Automated Software Engineering},
  pages={2378--2381},
  year={2024}
}

@inproceedings{tran2024chainsniper,
  title={Chainsniper: A machine learning approach for auditing cross-chain smart contracts},
  author={Tran, Tuan-Dung and Vo, Kiet Anh and Phan, Duy The and Tan, Cam Nguyen and Pham, Van-Hau},
  booktitle={Proceedings of the 2024 9th International Conference on Intelligent Information Technology},
  pages={223--230},
  year={2024}
}

@article{hu2025memory,
  title={Memory in the age of ai agents},
  author={Hu, Yuyang and Liu, Shichun and Yue, Yanwei and Zhang, Guibin and Liu, Boyang and Zhu, Fangyi and Lin, Jiahang and Guo, Honglin and Dou, Shihan and Xi, Zhiheng and others},
  journal={arXiv preprint arXiv:2512.13564},
  year={2025}
}

@article{zhang2025survey,
  title={A survey on the memory mechanism of large language model-based agents},
  author={Zhang, Zeyu and Dai, Quanyu and Bo, Xiaohe and Ma, Chen and Li, Rui and Chen, Xu and Zhu, Jieming and Dong, Zhenhua and Wen, Ji-Rong},
  journal={ACM Transactions on Information Systems},
  volume={43},
  number={6},
  pages={1--47},
  year={2025},
  publisher={ACM New York, NY}
}

@article{zhao2023depth,
  title={An in-depth survey of large language model-based artificial intelligence agents},
  author={Zhao, Pengyu and Jin, Zijian and Cheng, Ning},
  journal={arXiv preprint arXiv:2309.14365},
  year={2023}
}

@article{xie2024large,
  title={Large multimodal agents: A survey},
  author={Xie, Junlin and Chen, Zhihong and Zhang, Ruifei and Wan, Xiang and Li, Guanbin},
  journal={arXiv preprint arXiv:2402.15116},
  year={2024}
}

@article{huang2026rethinking,
  title={Rethinking Memory Mechanisms of Foundation Agents in the Second Half: A Survey},
  author={Huang, Wei-Chieh and Zhang, Weizhi and Liang, Yueqing and Bei, Yuanchen and Chen, Yankai and Feng, Tao and Pan, Xinyu and Tan, Zhen and Wang, Yu and Wei, Tianxin and others},
  journal={arXiv preprint arXiv:2602.06052},
  year={2026}
}

@misc{Nomad,

  
  title={Nomad Bridge Hack: Root Cause Analysis}, 
  howpublished = "\url{https://medium.com/nomad-xyz-blog/nomad-bridge-hack-root-cause-analysis-875ad2e5aacd}",
  note = "[Accessed 27-June-2026]",
}

@misc{Synapse,

  
  title={Synapse Exploit}, 
  howpublished = "\url{https://medium.com/synapse-protocol/11-06-2021-post-mortem-of-synapse-metapool-exploit-3003b4df4ef4}",
  note = "[Accessed 27-June-2026]",
}

@misc{statisticforOFCV,

  title={The statistic source for OFC vulnerability}, 
  howpublished = "\url{https://doi.org/10.5281/zenodo.21006162}",
  year = "2026",
  note = "[Accessed 27-June-2026]",
}

@article{wangpatterns,
  title={From Patterns to Precision: LLM-Guided Detection of Signature Verification Flaws in Smart Contracts},
  author={Wang, Huixin and Yan, Kailun and Diao, Wenrui},
  journal={IEEE International Conference on Software Analysis, Evolution and Reengineering (SANER)},
  pages={240--251},
  year={2026},
  organization={IEEE}
}

@inproceedings{sun2024gptscan,
  title={Gptscan: Detecting logic vulnerabilities in smart contracts by combining gpt with program analysis},
  author={Sun, Yuqiang and Wu, Daoyuan and Xue, Yue and Liu, Han and Wang, Haijun and Xu, Zhengzi and Xie, Xiaofei and Liu, Yang},
  booktitle={Proceedings of the IEEE/ACM 46th international conference on software engineering},
  pages={1--13},
  year={2024}
}

@misc{Thorchain,
  author = "{Sebastian Sinclair}",
  title={Blockchain Protocol Thorchain Suffers \$8M Hack}, 
  howpublished = "\url{https://www.coindesk.com/markets/2021/07/23/blockchain-protocol-thorchain-suffers-8m-hack/}",
  year = "2021",
  note = "[Accessed 20-Sep-2023]",
}

@misc{PolyNetworkexploit,
 
  title={PolyNetwork exploit}, 
  howpublished = "\url{https://en.wikipedia.org/wiki/Poly_Network_exploit}",
  year = "2022",
  note = "[Accessed 27-June-2026]",
}

@misc{Solidity,
  title={Solidity}, 
  howpublished = "\url{http://solidity.readthedocs.io/}",
  note= "[Accessed 20-Sep-2023]"
}

@inproceedings{feist2019Slither,
  title={Slither: a static analysis framework for smart contracts},
  author={Feist, Josselin and Grieco, Gustavo and Groce, Alex},
  booktitle={2019 IEEE/ACM 2nd international workshop on emerging trends in software engineering for blockchain (WETSEB)},
  pages={8--15},
  year={2019},
  organization={IEEE}
}

@article{chen2020defining,
  title={Defining smart contract defects on ethereum},
  author={Chen, Jiachi and Xia, Xin and Lo, David and Grundy, John and Luo, Xiapu and Chen, Ting},
  journal={IEEE Transactions on Software Engineering},
  volume={48},
  number={1},
  pages={327--345},
  year={2020},
  publisher={IEEE}
}

@inproceedings{liao2023smartstate,
  title={Smartstate: Detecting state-reverting vulnerabilities in smart contracts via fine-grained state-dependency analysis},
  author={Liao, Zeqin and Hao, Sicheng and Nan, Yuhong and Zheng, Zibin},
  booktitle={Proceedings of the 32nd ACM SIGSOFT International symposium on software testing and analysis},
  pages={980--991},
  year={2023}
}

@inproceedings{chen2024exploring,
  title={Exploring the security issues of real world assets (rwa)},
  author={Chen, Shijian and Jiang, Muhui and Luo, Xiapu},
  booktitle={Proceedings of the Workshop on Decentralized Finance and Security},
  pages={31--40},
  year={2024}
}

@inproceedings{guan2025security,
  title={Security perceptions of users in stablecoins: Advantages and risks within the cryptocurrency ecosystem},
  author={Guan, Maggie Yongqi and Yu, Yaman and Sharma, Tanusree and Huang, Molly Zhuangtong and Qin, Kaihua and Wang, Yang and Wang, Kanye Ye},
  booktitle={2025 IEEE Symposium on Security and Privacy (SP)},
  pages={2753--2771},
  year={2025},
  organization={IEEE}
}

@article{liao2024smartaxe,
  title={Smartaxe: Detecting cross-chain vulnerabilities in bridge smart contracts via fine-grained static analysis},
  author={Liao, Zeqin and Nan, Yuhong and Liang, Henglong and Hao, Sicheng and Zhai, Juan and Wu, Jiajing and Zheng, Zibin},
  journal={Proceedings of the ACM on Software Engineering},
  volume={1},
  number={FSE},
  pages={249--270},
  year={2024},
  publisher={ACM New York, NY, USA}
}

@inproceedings{luu2016making,
  title={Making smart contracts smarter},
  author={Luu, Loi and Chu, Duc-Hiep and Olickel, Hrishi and Saxena, Prateek and Hobor, Aquinas},
  booktitle={Proceedings of the 2016 ACM SIGSAC conference on computer and communications security},
  pages={254--269},
  year={2016}
}
